\documentstyle[12pt]{article}
\makeatletter
\newcommand{\singlespacing}{\let\CS=\@currsize\renewcommand{\baselinestretch}
{1.0}\tiny\CS}
\newcommand{\doublespacing}{\let\CS=\@currsize\renewcommand{\baselinestretch}
{1.5}\tiny\CS}
\begin{document}
\newcommand{\bd}{\begin{document}}
\newcommand{\ed}{\end{document}}
\newcommand{\bc}{\begin{center}}
\newcommand{\ec}{\end{center}}
\newcommand{\bfr}{\begin{flushright}}
\newcommand{\efr}{\end{flushright}}
\newcommand{\lt}{\left}
\newcommand{\rt}{\right}
\newcommand{\vs}{\vspace}
\newcommand{\hs}{\hspace}
\newcommand{\beq}{\begin{equation}}
\newcommand{\eeq}{\end{equation}}
\newcommand{\lb}{\linebreak}
\renewcommand{\pb}{\pagebreak}
\newcommand{\mb}{\makebox}
\newcommand{\fb}{\framebox}
\newcommand{\mc}{\multicolumn}
\newcommand{\ben}{\begin{enumerate}}
\newcommand{\een}{\end{enumerate}}
\newcommand{\bit}{\begin{itemize}}
\newcommand{\eit}{\end{itemize}}
\newcommand{\ol}{\overline}
\newcommand{\un}{\underline}
\newcommand{\lefq}{\lefteqn}
\newcommand{\ba}{\begin{array}}
\newcommand{\ea}{\end{array}}
\newcommand{\beqa}{\begin{eqnarray}}
\newcommand{\eeqa}{\end{eqnarray}}
\newcommand{\beqas}{\begin{eqnarray*}}
\newcommand{\eeqas}{\end{eqnarray*}}
\newcommand{\bfg}{\begin{figure}}
\newcommand{\efg}{\end{figure}}
\newcommand{\bds}{\begin{displaymath}}
\newcommand{\eds}{\end{displaymath}}
\newcommand{\btb}{\begin{tabbing}}
\newcommand{\etb}{\end{tabbing}}
\newcommand{\para}{\parallel}
\newcommand{\pad}{\partial}
\newcommand{\nn}{\nonumber}
\newcommand{\la}{\leftarrow}
\newcommand{\ra}{\rightarrow}
\newcommand{\lgla}{\longleftarrow}
\newcommand{\lgra}{\longrightarrow}
\newcommand{\La}{\Leftarrow}
\newcommand{\Ra}{\Rightarrow}
\newcommand{\Lra}{\Leftrightarrow}
\newcommand{\Lgla}{\Longleftarrow}
\newcommand{\Lgra}{\Longrightarrow}
\newcommand{\bm}{\boldmath}
\newcommand{\lan}{\langle}
\newcommand{\ran}{\rangle}
\renewcommand{\a}{\alpha}
\renewcommand{\b}{\beta}
\newcommand{\g}{\gamma}
\newcommand{\G}{\Gamma}
\renewcommand{\d}{\delta}
\newcommand{\eps}{\epsilon}
\newcommand{\th}{\theta}
\newcommand{\Th}{\Theta}
\newcommand{\s}{\sigma}
\newcommand{\lam}{\lambda}
\newcommand{\D}{\Delta}
\newcommand{\vare}{\varepsilon}
\newcommand{\pr}{\prime}
\newcommand{\ro}{\rho}
\newcommand{\nab}{\nabla}
\newcommand{\m}{\mu}
\newcommand{\n}{\nu}
\newcommand{\Sg}{\Sigma}
\newcommand{\p}{\pi}
\newcommand{\R}{I\!\!R}
\newcommand{\om}{\omega}
\newcommand{\Om}{\Omega}
\newcommand{\ze}{\zeta}
\newcommand{\vart}{\vartheta}
\newcommand{\tri}{\triangle}
\newcommand{\f}{\frac}
\newcommand{\iny}{\infty}
\newcommand{\pro}{\propto}

\bc

{\huge \bf Three Dimensional Confinement : }

\vs{.5cm}

{\huge \bf WKB Revisited}

\ec

\vspace{1cm}

\bc
{\Large {\it {\bf Anjana Sinha $^*$}} }
\ec


\bc
{\large {\it Dept. of Applied Mathematics}}\\ 
{\large {\it Calcutta University} }\\ 
{\large {\it 92, A.P.C. Road} }\\
{\large {\it Kolkata - 700 009}}\\ 
{\large {\it INDIA}}\\
\ec


\vs{6cm}

\noindent
$^*$ ~ e-mail : a.sinha@cucc.ernet.in \\
\hspace*{2.2cm}{anjana23@rediffmail.com}


\thispagestyle{empty}

\setlength{\baselineskip}{19.5pt}

\pb

\bc

{\bf \large \un{Abstract}}
\ec

An alternate formalism is developed to determine the energy eigenvalues of 
quantum mechanical systems, confined within a rigid impenetrable 
spherical box of radius $r_0$, in the framework of 
Wentzel-Kramers-Brillouin (WKB) approximation.
Instead of considering the Langer correction for the centrifugal term,
the approach adopted here is that of Hainz \& Grabert : 
The centrifugal term is expanded perturbatively (in powers 
of $\hbar$), decomposing it into 2 terms --- the classical centrifugal
potential and a quantum correction.
Hainz and Grabert found that this method reproduced the exact energies of the 
hydrogen atom, to the first order in $ \hbar $, with 
all higher order corrections vanishing. 
In the present study, this formalism is extended to the 
case of radial potentials under hard wall confinement, to
check whether the same argument holds good for such confined systems as well. 
As expicit examples, 3 widely known potentials are studied, which are 
of considerable importance in the theoretical treatment of 
various atomic phenomena involving atomic transitions, viz., 
the 3-dimensional Harmonic oscillator, the hydrogen atom 
and the Hulthen potential.

\vs{5cm}

\noindent
key words : WKB approximation, 3-dimensional spatial confinement,
radial potentials, perturbative expansion, centrifugal term,
harmonic oscillator, hydrogen atom, Hulth\'{e}n potential

\pb

\section*{1. Introduction }

Spatial confinement of electrons in artificial nanostructures, on a scale
comparable to their de Broglie wavelength, is a much talked about 
subject for the past decade or so [1-7]. Such {\it artificial atoms} as they
are called because of their quantized energies, undergo 
radical changes in terms of both physical and chemical properties,
because of their extremely small spatial dimensions,
making them very useful in
the study of atomic and molecular phenomena [8]. However, 
the problem with these so-called quantum wells, quantum wires 
and quantum dots is that their exact analytical treatment is not possible in
most of the cases. Consequently, various approximation methods --- the
variational approach, the shifted 1/N expansion technique, the modified Airy 
function (MAF) method, the supersymmetric version of the same (SMAF),
the Wentzel-Kramers-Brillouin (WKB) approximation, its supersymmetric 
version (SWKB), etc. --- come into the picture [9-18]. The other option
is to go for a numerical solution. Of the several approximation methods, 
the semiclassical WKB
approximation technique is a very effective tool,
yielding fairly accurate results in various 
quantum mechanical problems, where exact solutions are unknown or 
difficult to find out. This method gives a good estimate of 
both the energy eigenvalues as well as eigenfunctions, with the exception of
the region near the classical turning points. Though this approach 
works well for one-dimensional problems, practical use shows that 
the standard leading order WKB approximation always 
reproduces the exact spectrum for the solvable spherically symmetric
potentials $V(r)$ if the centrifugal term 
$$ V_C (r) = l(l+1) \hbar ^2 /2r^2 $$
is replaced by the Langer correction term [19] 
$$ V_L (r) = (l+1/2)^2 \hbar ^2 /2r^2 $$
This modification also justifies 
the WKB expansion of singular potentials like that of the Coulomb potential, 
near the origin. However, some authors have attempted to get rid of this
Langer modification (LM), based on non-linear transformations [20] and 
supersymmetry [21]. Though their approach yielded results which are superior to
those with LM, Coulomb type problems did not fare well. Hainz and Grabert [22]
challenged this common belief and put forward a new method to deal with
centrifugal terms in the WKB approximation. 

Since the semiclassical WKB approximation proceeds as a perturbation in powers
of $ \hbar $, it was argued in ref. [22] that within this expansion, the
centrifugal term can be decomposed as
\beq
V_c (r) = \f{l(l+1) \hbar ^2}{2m r^2} 
= \frac{L_0 ^2}{2mr^2} + \frac{\hbar L_0}{2mr^2} 
\eeq
with $ L_0 = \hbar l $. The first term is the classical centrifugal term, 
while the second term is a quantum correction. 
Thus the quantum correction can be treated as a perturbation 
and expanded accordingly. Proceeding along these lines,
Hainz and Grabert [22] found that the semiclassical energy eigenvalues for the
hydrogen atom turned out to be exact to the first order in $ \hbar $, with 
all higher order corrections vanishing. The aim of the present work is to
check whether the same argument holds good for confined systems as well. The
formalism of ref. [22] is extended to the case of radial potentials confined
within rigid impenetrable spheres of radius $r_0$. As expicit examples, 3
widely known cases are studied, viz., \\
~~(i) ~~ the 3-dimensional Harmonic oscillator (HO) \\
~(ii) ~~ the hydrogen atom \\
(iii) ~~ the Hulthen potential \\
These potentials are of considerable importance in theoretical treatment of
various atomic phenomena involving atomic transitions. 
It is observed that this formalism 
gives a better estimate of the energy spectrum
than the case with Langer modification, even in case of spatially confined
systems. 

The organisation of the paper is as follows. In Section II, the WKB
formalism is extended to include quantum 
mechanical systems confined radially, expanding the 
centrifugal term perturbatively in powers of $ \hbar $. In Section III,
the approach is applied explicitly to 3 physically relevant cases, viz.,
the 3-dimensional harmonic oscillator, the hydrogen atom, and the Hulthen
potential. Section IV is kept for conclusions and discussions.

\vs{2cm}

\section*{2. Theory}

The starting point of the study is the three-dimensional 
Schr\"{o}dinger equation for a radial potential $V(r)$
\beq
\frac{d^2}{dr^2} ~~ \psi(r) ~+~ 
\f{2m}{{\hbar}^{~2}} \lt [ E ~-~ V(r) ~-~ \f{l(l+1) \hbar ^2}{2mr^2} \rt ] 
~ \psi(r) ~=~ 0 
\eeq
where $ l(l+1) {\hbar}^2 $ represents the eigen values of the square
of the angular momentum operator $ L^2 $ and $m$ is the mass of the 
particle. It is worth noting here that the WKB approximation 
can be applied only when the de Broglie wavelength $ \lam = h/p $
~ $(h=2 \pi \hbar) $ is changing slowly.
With the help of $(1)$ the radial Schr\"{o}dinger equation $(2)$
can be cast in the form
\beq
\f{d^2}{dr^2} ~~ \psi(r) ~+~ 
\f{2m}{{\hbar}^{~2}} \lt [ E ~-~ V_{eff}(r) ~-~ \hbar \f{L_0}{2mr^2} \rt ] 
~ \psi(r) ~=~ 0 
\eeq
where 
\beq
V_{eff} (r) = V(r) + \f{L_0 ^2}{2mr^2} 
\eeq
In order for the physical system to have a stable bound state (discrete 
spectrum) it must have two classical turning points $r_1$ and $r_2$.
This gives rise to 3 regions given by \\
region ~~I ~:~ $ 0~~<~r~<~r_1 $ ~:~  $ V_1 ~>~ E $  \\
region ~II ~:~ $r_1 ~<~ r ~<~ r_2 $ ~:~  $ E ~>~ V_1 $ \\
region III ~:~ $ r ~~>~~ r_2 $ ~~~~~ ~:~  $ V_1 ~>~ E $ ,\\
where
\beq
V_1 (r) =  V_{eff} (r)  + \hbar \f{L_0}{2mr^2} 
\eeq
If one defines $\Gamma (r) $ and $ \kappa (r) $, 
with $ ( \kappa ^2 = - \Gamma ^2 ) $  by
\beq
\Gamma (r) = \sqrt{\f{2m}{\hbar ^2}\lt \{ \lt (E - V_{eff} \rt ) - \hbar
\f{L_0}{2m r^2} \rt \} }
\eeq
\beq
\kappa (r) = \sqrt{\f{2m}{\hbar ^2}\lt \{ \lt ( V_{eff} - E \rt ) + \hbar
\f{L_0}{2m r^2} \rt \} }
\eeq
then $(3)$ reduces to
\beq
\lt \{ \f{d^2}{dr^2} - \kappa ^2 (r) \rt \} \psi = 0 ~~~~~~~~~~ \rm in \ \
regions \ \ I \ \ and \ \ III 
\eeq
\beq
\lt \{ \f{d^2}{dr^2} + \Gamma ^2 (r) \rt \} \psi = 0 ~~~~~~~~~~ \rm in \ \
\ \ region \ \ II 
\eeq
Expanding in powers of $\hbar$, and keeping terms to the first order in $
\hbar $, one can write $ \Gamma (r)$ and $ \kappa (r)$  as 
\beq
\Gamma (r) \simeq \Gamma _0 (r) - \f {L_0}{2 \hbar \Gamma _0 r^2 }
\eeq
\beq
\kappa (r) \simeq \kappa _0 (r) + \f {L_0}{2 \hbar \kappa _0 r^2 }
\eeq
where
\beq
\Gamma _0 (r) = \sqrt{ \f{2m}{\hbar ^2} \lt \{ E - V_{eff} \rt \} }
\eeq
\beq
\kappa _0 (r) = \sqrt{ \f{2m}{\hbar ^2} \lt \{ V_{eff} - E \rt \} }
\eeq
so that 
\beq
\f{1}{\sqrt{\Gamma (r)}} \simeq \f{1}{\sqrt{\Gamma _0 (r)}} 
\lt \{ 1 + \f{L_0}{4 \hbar \Gamma _0 ^2 r^2 } \rt \}
\eeq
\beq
\f{1}{\sqrt{\kappa (r)}} \simeq \f{1}{\sqrt{\kappa _0 (r)}} 
\lt \{ 1 - \f{L_0}{4 \hbar \kappa _0 ^2 r^2 } \rt \}
\eeq
The conventional WKB ansatz is assumed for the wave function
\beq 
\psi(r) = exp \lt [ \f{i}{\hbar} \Sg (-i \hbar)^k S_k (r) \rt] 
\eeq
Substituting 
\beq
y_k (r) = \f{\partial S_k (r)}{\partial r} 
\eeq
and expanding them in powers of $\hbar$, one obtains the set of relations
\beq
y_0 = \pm \sqrt{2m \lt ( E-V_{eff} (r) \rt ) } 
\eeq
\beq
y_1 = - \f{1}{2y_0} \lt ( y_0 ^{\prime} + i \f{L_0}{r^2} \rt ) 
\eeq
\beq
y_{2m} = - \f{1}{2y_0} \lt\{ y_m ^2 + y_{2m - 1 } ^{\pr} + 2 \sum _{k=1}
^{2m-2} ~ y_{2m-k} ~ y_k \rt\}
\eeq
\beq
y_{2m+1} = - \f{1}{2y_0} \lt\{ y_{2m} ^{\pr}  + 2 \sum _{k=1}
^{2m-1} ~y_{2m+1-k} ~y_k \rt\}
\eeq
In the above relationships, prime denotes differentiation with respect to $r$.
Thus $y_0$ turns out to be the classical momentum, and \\
$ y_0 = \pm i \kappa _0 \hbar $ in regions I and III, \\
$ y_0 = \Gamma _0 \hbar $ in region II.\\
So, the wave function is a linear combination of the form
\beq
\psi(r) = \Sg c^{\pm} exp \lt [ \f{i}{\hbar} \int 
dr ~ y^{\pm} (r) \rt] 
\eeq
where  
\beq
y(r) = \Sg (-i \hbar)^k y_k (r) 
\eeq
with derivatives (to the first order in $\hbar$)
\beq
\f{d \psi}{dr} = \lt ( \f{i}{\hbar} S_0 ^{~\pr} + S_1 ^{~\pr} \rt ) \psi
\eeq
\beq
\f{d^2 \psi}{dr^2} = \lt \{ -\f{S_0 ^{~\pr ~2}}{\hbar ^2} +
\f{i}{\hbar} \lt ( 2S_0 ^{~\pr} S_1 ^{~\pr} + S_0 ^{~\pr ~\pr} \rt ) 
+ S_1 ^{~\pr ~\pr} \rt \} \psi
\eeq
This gives the complete solution to the Schr\"{o}dinger equation 
( to the first order in $\hbar$ ) as
\beq
\psi = \f{1}{\sqrt{y_0}} ~ exp \lt \{ \f{i}{\hbar} ~ \int y_0 ~dr - \f{iL_0}{2}
~ \int \f{dr}{y_0 r^2} \rt \}
\eeq
Since the radial wave function must vanish at $r=0$, the only allowed solution
in region I is
\beq
\psi _I = \f{A}{\sqrt{\kappa _0}} exp \lt \{ - \int_r ^{r_1} \kappa _0 dr - 
\f{L_0}{2 \hbar} \int_r ^{r_1} \f{dr}{\kappa _0 r^2} \rt \}
\eeq
Now we shall make use of the conventional connection formulae for WKB
approximation at the turning point $ r_1 $, [23] {\it viz.},
\beq
\f{1}{\sqrt{\kappa (r)} } ~ exp \lt ( - \int _r ^{r_1} \kappa (r) ~ dr \rt )
~=~ \f{2}{\sqrt{\Gamma (r) }} ~ sin \lt \{ \int _{r_1} ^r \Gamma (r) ~ dr ~+~
\f{\pi}{4} \rt \}
\eeq
\beq
\f{1}{\sqrt{\kappa (r)} } ~ exp \lt ( ~ \int _r ^{r_1} \kappa (r) ~ dr \rt )
~=~ \f{1}{\sqrt{\Gamma (r) }} ~ cos \lt \{ \int _{r_1} ^r \Gamma (r) ~ dr ~+~
\f{\pi}{4} \rt \}
\eeq 
Expanding $ \Gamma (r) $ and $ \kappa (r) $ in terms of 
$ \Gamma _0 (r) $ and $ \kappa _0 (r) $, and keeping terms upto 
the first order in $ \hbar $ only, the connection formulae can be cast in the
form 
$$ \f{1}{\sqrt{\kappa _0 (r)} } ~ exp \lt ( - \int _r ^{r_1} 
\kappa _0(r) ~ dr ~-~ \f{L_0}{2 \hbar} ~ \int _r ^{r_1} 
\f{dr}{ \kappa _0 (r) r^2 }  \rt ) $$
\beq
~\simeq~ \f{2}{\sqrt{\Gamma _0 (r) }} ~ sin \lt \{ 
\int _{r_1} ^r \Gamma _0(r) ~ dr  
~-~ \f{L_0}{2 \hbar} ~ \int _{r_1} ^r ~ \f{dr}{ \Gamma _0 (r) r^2 }
~+~ \f{\pi}{4} \rt \}
\eeq
$$ \f{1}{\sqrt{\kappa _0(r)} } ~ exp \lt ( ~ \int _r ^{r_1} 
\kappa _0(r) ~ dr  ~+~ \f{L_0}{2 \hbar} ~ \int _r ^{r_1} 
\f{dr}{ \kappa _0 (r) r^2 } \rt ) $$
\beq
~\simeq~ \f{1}{\sqrt{\Gamma _0(r) }} ~ cos \lt \{ 
\int _{r_1} ^r \Gamma _0 (r) ~ dr 
~-~ \f{L_0}{2 \hbar} ~ \int _{r_1} ^r ~ \f{dr}{ \Gamma _0 (r) r^2 }
~+~ \f{\pi}{4} \rt \}
\eeq 
The solutions to the Schr\"{o}dinger 
equation in regions II and III are obtained by 
matching the WKB solutions on either side of the turning points $r_1$ and
$r_2$, with the help of the connection formulae $(30)$ and $(31)$.
Thus the solution in region II comes out to be
\beq
\psi _{II} ~=~ \f{2A}{\sqrt{ \Gamma _0 (r) }} ~ sin  \lt \{ \int _{r_1} ^{~r}
\Gamma _0 (r)  dr ~-~ \f{L_0}{2 \hbar} \int _{r_1} ^{~r} \f{dr}{ \Gamma _0 (r)
r^2 } ~+~ \f{\pi}{4} \rt \}
\eeq
To obtain the solution in region III, $\psi _{II} $ is written as
\beq
\ba{l}
\int _{r_1} ^{r}
\Gamma _0 (r)  dr - \f{L_0}{2 \hbar} \int _{r_1} ^{r} \f{dr}{ \Gamma _0 (r)
r^2 } + \f{\pi}{4}  \\
\\
= \int _{r_1} ^{r_2} \Gamma _0 (r)  dr 
- \int _{r} ^{r_2} \Gamma _0 (r)  dr 
- \f{L_0}{2 \hbar} \int _{r_1} ^{r_2} \f{dr}{ \Gamma _0 (r) r^2 } 
+ \f{L_0}{2 \hbar} \int _{r} ^{r_2} \f{dr}{ \Gamma _0 (r) r^2 } 
+ \f{\pi}{4}  \\
\\
= \lt( \f{\pi}{2} + \int _{r_1} ^{r_2} \Gamma _0 (r)  dr 
- \f{L_0}{2 \hbar} \int _{r_1} ^{r_2} \f{dr}{ \Gamma _0 (r) r^2 } \rt)
- \lt( \int _{r} ^{r_2} \Gamma _0 (r)  dr 
- \f{L_0}{2 \hbar} \int _{r} ^{r_2} \f{dr}{ \Gamma _0 (r) r^2 } 
+ \f{\pi}{4} \rt) \\
\\
= \lt( \f{\pi}{2} + \theta \rt) - B
\ea
\eeq
where 
\beq
\theta ~=~ \int _{r_1} ^{~r_2} \Gamma _0 (r) ~ dr ~-~ \f{L_0}{2 \hbar}
 \int _{r_1} ^{~r_2} \f{dr}{ \Gamma _0 r^2}
\eeq
and
\beq
B ~=~  \int _{r} ^{r_2} \Gamma _0 (r) ~ dr ~-~ \f{L_0}{2 \hbar}
 \int _{r} ^{r_2} \f{dr}{ \Gamma _0 r^2} + \f{\pi}{4}
\eeq
Substituting $(33)$ into $(32)$ and simplifying, 
\beq
\psi _{II} ~=~ \f{2A}{\sqrt{ \Gamma _0 (r) }} \lt( cos ~ \theta ~ cos B
+ sin ~ \theta ~ sin ~B \rt)
\eeq
Using the connection formula at the turning point $r_2$, the solution in
region III is obtained as
\beq
\psi _{III} ~=~ \f{2A}{\sqrt{ \kappa _0 (r) }} ~cos~ \theta ~ exp(\sigma (r))
~+~ \f{A}{\sqrt{\kappa _0 (r) }} ~ sin ~ \theta ~ exp ( - \sigma (r) )
\eeq
where
\beq
\sigma (r) ~=~  \int _{r_2} ^{r} \kappa _0 (r) ~ dr ~-~ \f{L_0}{2 \hbar}
 \int _{r_2} ^{r} \f{dr}{ \kappa _0 r^2}
\eeq
Now, the WKB quantization rule is obtained by the constraint
$ \psi ( r = r_0) ~=~ 0 $, 
where $r_0$ is the radius of the confining spherical box. 
2 cases arise depending on the size of confinement, i.e. \\
~i) there is a single turning point within the box ( $r_1 < r_0 < r_2$ )\\
ii) both the turning points are within the box ($ r_0 > r_2 $)\\
thus yielding 2 different quantization rules.

\vs{1cm}

\noindent
{\bf \un{1) Extremely small confinement : }}\\
The size of the box is so small that it admits only a single turning point, 
i.e., $ r_1 ~<~ r_0 ~<~ r_2 $ \\
This modifies the boundary condition to : $ ~~~ \psi _{II} (r_0) ~=~ 0 $. \\
Hence the WKB quantization condition reads
\beq
\lam _1 ~-~ \lam _2 ~=~ \lt ( n + \f{3}{4} \rt ) , \ \ \ \ \ \ n=0,1,2,.....
\eeq
with
\beq
\lam _1 ~=~  \int _{r_1} ^{~r_0} \Gamma _0 (r)  dr 
\eeq
\beq
\lam _2 ~=~ \f{L_0}{2 \hbar} \int _{r_1} ^{~r_0} \f{dr}{ \Gamma _0 (r) r^2 }
\eeq

\vs{1cm}

\noindent
{\bf \un{2) The confining box is not so small :}} \\
The size of the box is such that both the classical turning points lie within
it, i.e., $r_0 ~>~ r_2 $ .\\
Hence, in this case the solution of the Schr\"{o}dinger equation must obey the
boundary condition : $ ~~~ \psi _{III} (r_0) ~=~ 0 $, \\
yielding the WKB quantization rule 
\beq
2~cos ~ \theta . ~ exp \lt ( \sigma (r_0) \rt ) ~+~ sin ~ \theta . ~ exp \lt 
( - \sigma (r_0) \rt ) ~=~ 0
\eeq

\vs{.5cm}

\noindent
These quantization rules $(39)$ and $(42)$ enable one to determine the energy
spectrum of any spatially confined, radial potential, in the
framework of WKB approximation.

\vs{2cm}

\section*{3. Calculations}

In this section, the WKB rules so developed are used to determine the energy
spectrum of 3 explicit potentials \\
~~(i) ~~ the 3-dimensional Harmonic oscillator \\
~(ii) ~~ the hydrogen atom \\
(iii) ~~ the Hulthen potential \\
each confined within an impenetrable spherical box of radius $r_0$. 
All the 3 potentials are of tremendous importance in a 
variety of physical problems, and
have been studied widely. 
Units used are $ \hbar ~=~ m ~=~ 1 $ so that $ L_0 = l $.

\vs{1cm}

\noindent
{\bf \un{3-dimensional Harmonic Oscillator}}

\beq
V(r) ~=~ \f{r^2}{2} 
\eeq
Hence the relationship $ E - V_{eff} = 0 $
gives the classical turning points at
\beq
r_1 ~=~  \lt \{ E - \sqrt{ E^2 - l^2} \rt \} ^{1/2} 
\eeq
\beq
r_2 ~=~  \lt \{ E + \sqrt{ E^2 - l^2} \rt \} ^{1/2} 
\eeq
Proceeding along the formalism developed above, 
the energy levels of the enclosed 3-dim. 
harmonic oscillator are computed for various values of the confining parameter
$r_0$, with the help of the mathematical relationships in 
ref. [24]. 
The results are presented in Table I, comparing the energies so obtained 
(with no modification of the centrifugal term) denoted by E, 
with those from the conventional WKB quantization rules for
3-dimensional confinement (with Langer modification), E(WKB)  [18], the direct
variational method, E(var) [9], 
and exact numerical values, E(exact) [9]
for the enclosed 3-dim. harmonic oscillator. 

\vs{1cm} 

\noindent
{\bf Table I : Enclosed 3-dim. Harmonic Oscillator } \\
($ n=n_r + l + 1 : n_r = 0, l=1$)

\begin{tabular}{|c|c|c|c|c|} \hline
$r_0$ & E & E(WKB)  & E(var)  & E(exact) \\ \hline
1.0   & 10.2876   & 10.2643 & 10.3188 & 10.2822  \\ \hline
1.5   & ~4.9068   & ~4.9084 & ~4.9169 & ~4.9036  \\ \hline
2.0   & ~3.3081$^*$   & ~3.2490 & ~3.2514 & ~3.2469  \\ \hline
2.5   & ~2.6835   & ~2.7079$^*$ & ~2.6901 & ~2.6881  \\ \hline
3.0   & ~2.5313   & ~2.5310 & ~2.5337 & ~2.5313  \\ \hline
4.0   & ~2.5001   & ~2.5001 & ~2.5015 & ~2.5001  \\ \hline
5.0   & ~2.5000   & ~2.5000 & ~2.5012 & ~2.5000  \\ \hline
\end{tabular}

\vs{.5cm}
$*$ In these cases the size of the rigid spherical box is such that the wall
is close to the turning point, where the WKB approximation is not expected to
give good results.

\vs{1cm}

\noindent
{\bf \un{Hydrogen Atom}}

\noindent
The well-known Coulomb potential 
\beq
V(r) ~=~ -~ \frac{1}{r} 
\eeq
is known to possess negative energies. However, spatial confinement alters
this scenario. For extremely small confinement, 
the system is no longer bound. $ E ~>~ 0 $,
and $ E - V_{eff} = 0 $ gives a single turning point at
\beq
r_t ~=~ \f{ \sqrt{ 1 + 2E l^2 } ~-~ 1}{2E}
\eeq
For all practical purposes, $r_t$ is very small, and the 
eigenenergies of the enclosed system are obtained from the
relationship (39)
\beq
\int _{r_t} ^{r_0}  \sqrt{ 2 \lt ( E -V_{eff} \right )} ~ dr 
~-~ \f{l}{2} ~ \int _{r_t} ^{r_0} ~ \f{dr}{\sqrt{ 2 
\lt ( E -V_{eff} \rt ) }~ r^2 } 
~=~ \lt ( n ~+~ \f{3}{4} \rt ) \pi
\eeq
with 
\beq
V_{eff} ~=~ - \f{1}{r} ~+~ \f{l^2}{2 r^2}
\eeq

\vspace{.5cm}

\noindent
However, for bound energies $( E~<~0 )$, 2 cases may arise, 
depending on the size of the confining box : \\
Either there is only one turning point inside the box 
$( r_1 ~<~ r_0 ~<~ r_2 )$, 
or confining wall encloses both the turning points  $( r_0 ~>~ r_2 )$.
Let $ E ~=~ -e $ and $ V_{eff} ~=~ - v_{eff} $.
The expression $ E - V_{eff}  = 0 $ gives the roots at
\beq
r_1 ~=~ \f{1}{2e} ~ \lt \{ 1 ~-~ \sqrt{ 1 ~-~ 2el^2 } \rt \}
\eeq
\beq
r_2 ~=~ \f{1}{2e} ~ \lt \{ 1 ~+~ \sqrt{ 1~-~ 2el^2 } \rt \}
\eeq
The energy spectrum of the boxed-in hydrogen atom is determined 
with the help of the formalism developed above.  
The results are computed and presented in tabular form,
for the $2p$ (Table II)  and  $3d$ (Table III) states. 
The energy eigenvalues calculated in this study, E, are compared 
with those obtained by the conventional (with LM) confined 
WKB approximation developed earlier  E(WKB) [18],  the 
direct variational method of Marin and Cruz, E(var) [9],
Varshni's modification of Marin-Cruz approach, E(Varshni) [25],
and the exact numerical values, E(exact) [25,26]. 

\pb

\noindent
{\bf Table II : Enclosed Hydrogen Atom - $2p$ state} :
($ n = n_r + l + 1 : n_r = 0, l = 1 $) \\
\begin{tabular}{|c|c|c|c|c|c|} \hline
 $r_0$ &   E &  E(WKB)  & E(var)    & E(Varshni) & E(exact)   \\ \hline
~0.6   & 49.8448  & 49.3997     & 50.401    & 49.935     & 49.874    \\ \hline
~0.8   & 26.9179  & 26.5586     & 27.155    & 26.910     & 26.879    \\ \hline
~1.0   & 16.5063  & 16.2590     & 16.611    & 16.464     & 16.446    \\ \hline
~1.2   & 10.8828  & 10.7653     & 10.999    & 10.905     & 10.893    \\ \hline
~1.4   & ~7.6209  & ~7.4379     & ~7.6857   & ~7.6214    & ~7.6138    \\ \hline
~1.6   & ~5.5112  & ~5.3928     & ~5.5801   & ~5.5347    & ~5.5295    \\ \hline
~1.8   & ~4.1512$^*$  & ~4.0693     & ~4.1675   & ~4.1345    & ~4.1308    \\ \hline
~2.0   & ~3.1513  & ~3.1010     & ~3.1791   & ~3.1547    & ~3.1520    \\ \hline
~2.2   & ~2.4469  & ~2.4013     & ~2.4641   & ~2.4458    & ~2.4438    \\ \hline
~2.4   & ~1.9224  & ~1.8815     & ~1.9326   & ~1.9187    & ~1.9173    \\ \hline
~2.8   & ~1.2129  & ~1.1807     & ~1.2157   & ~1.2075    & ~1.2068   \\ \hline
~3.0   & ~0.9684  & ~0.9420     & ~0.9694   & ~0.9631    & ~0.9625   \\ \hline
~3.5   & ~0.5466  & ~0.5371     & ~0.5459   & ~0.5427    & ~0.5424   \\ \hline
~4.0   & ~0.2894  & ~0.2771     & ~0.2888   & ~0.2872    & ~0.2871   \\ \hline
~5.0   & ~0.0154  & ~0.0135     & ~0.0155   & ~0.0152    & ~0.0152   \\ \hline
~7.0   & -0.1687$^*$  & -0.1666     & -0.1748   & -0.1748    & -0.1749   \\ \hline
10.0   & -0.2269  & -0.2256     & -0.2369   &            & -0.2377   \\ \hline
14.0   & -0.2487  & -0.2484     & -0.2484   &            & -0.2491   \\ \hline
\end{tabular}

\vs{.5cm}

\noindent
{\bf Table III : Enclosed Hydrogen Atom - $3d$ state} :
($ n = n_r + l + 1 : n_r = 0, l = 2 $) \\
\begin{tabular}{|c|c|c|c|c|c|} \hline
$r_0$ & E & E(WKB)  & E(var)  & E(Varshni) & E(exact)   \\ \hline
~1.0   & 29.8203  & 29.7306  & 30.234  & 29.979     & 29.935     \\ \hline
~1.5   & 12.5321  & 12.4895  & 12.692  & 12.587     & 12.570    \\ \hline
~2.0   & ~6.6415  & ~6.6064  & ~6.7182 & ~6.6640    & ~6.6550   \\ \hline
~2.5   & ~3.9882  & ~3.9658  & ~4.0288 & ~3.9970    & ~3.9920   \\ \hline
~3.0   & ~2.5863  & ~2.5593  & ~2.6088 & ~2.5887    & ~2.5856   \\ \hline
~4.0   & ~1.2467  & ~1.2379  & ~1.2532 & ~1.2440    & ~1.2427   \\ \hline
~5.0   & ~0.6581  & ~0.6489  & ~0.6634 & ~0.6588    & ~0.6582   \\ \hline
~6.0   & ~0.3617  & ~0.3550  & ~0.3634 & ~0.3609    & ~0.3607   \\ \hline
~7.0   & ~0.1926  & ~0.1890  & ~0.1945 & ~0.1933    & ~0.1932   \\ \hline
~8.0   & ~0.0919  & ~0.0897  & ~0.0928 & ~0.0922    & ~0.0921   \\ \hline
10.0   & -0.0140  & -0.0156  & -0.0141 & -0.0142    & -0.0142   \\ \hline
12.0   & -0.0625  & -0.0626  & -0.0625 & -0.0625    & -0.0625   \\ \hline
14.0   & -0.0862  & -0.0860  & -0.0862 &            & -0.0862   \\ \hline
16.0   & -0.0939$^*$  & -0.0928 & -0.0982 &         & -0.0984   \\ \hline
20.0   & -0.1079  & -0.1077  & -0.1076 &            & -1.1079   \\ \hline  
\end{tabular}

\vspace{2cm}

\noindent
{\bf \un{Confined Hulth\'{e}n potential :}}

Screened Coulomb potentials  are of tremendous
importance in atomic phenomena. The particular case studied here is
the confined Hulth\'{e}n potential, given by
\beq
V(r) ~=~ V_0 ~  \f{e^{- \delta r}}{ 1~-~ e^{- \delta r}}
\eeq
where $ V_0 ~=~ -Z \delta $, with $Z$ the atomic number and $ \delta $ 
the screening parameter. 
Taking $Z = 1$,
\beq
V_{eff} (r) ~=~ V(r) ~+~ \f{l^2}{2r^2}
\eeq
Once again the energy eigenvalues are computed for 
different values of the confining
radius $r_0$ and screening parameter $ \delta $, and the results presented in
tabular form, for ease of comparison with 
those obtained from other approximation methods, viz., the shifted $1/N$
expansion method E(1/N) [15], and exact numerical energies E(exact) [15].

\vs{.5cm}

\noindent
{\bf Table IV : Confined Hulth\'{e}n potential } : 
($ \delta = 0.1, n=n_r+l+1$ ) \\
\begin{tabular}{|c|c|c|c|c|c|c|} \hline
$r_0$ & state & $n_r $ & $l$ & E & E(exact)  & E(1/N)     \\ \hline 
~6 & 2p & 0 & 1 & -0.00782 & -0.00865 & -0.00294 \\ \hline
~7 & 2p & 0 & 1 & -0.03976 & -0.04069 & -0.03324 \\ \hline
~8 & 2p & 0 & 1 & -0.05510 & -0.05783 & -0.05293 \\ \hline
~9 & 2p & 0 & 1 & -0.06612 & -0.06728 & -0.06389 \\ \hline
10 & 2p & 0 & 1 & -0.07196 & -0.07257 & -0.07008 \\ \hline
25 & 2p & 0 & 1 & -0.07921 & -0.07918 & -0.07920 \\
   & 3p & 1 & 1 & -0.01384 & -0.01475 & -0.01295 \\
   & 3d & 0 & 2 & -0.01381 & -0.01390 & -0.01332 \\ \hline
50 & 2p & 0 & 1 & -0.07920 & -0.07918 & -0.07920 \\
   & 3p & 1 & 1 & -0.01598 & -0.01605 & -0.01578 \\
   & 3d & 0 & 2 & -0.01450 & -0.01448 & -0.01450 \\ \hline
\end{tabular}

\vs{1cm}

\noindent
{\bf Table V : Confined Hulth\'{e}n potential } :
($ \delta = 0.2, n=n_r+l+1$ ) \\
\begin{tabular}{|c|c|c|c|c|c|c|} \hline
$r_0$ & state & $n_r $ & $l$ & E & E(exact)  & E(1/N)     \\ \hline 
~8 & 2p & 0 & 1 & -0.01607 & -0.01731 & -0.01242 \\ \hline
~9 & 2p & 0 & 1 & -0.02612 & -0.02749 & -0.02428 \\ \hline
10 & 2p & 0 & 1 & -0.03389 & -0.03339 & -0.03118 \\ \hline
25 & 2p & 0 & 1 & -0.04192 & -0.04188 & -0.04199 \\ \hline
50 & 2p & 0 & 1 & -0.04191 & -0.04189 & -0.04196 \\ \hline
\end{tabular}

\vs{1cm}

\section*{4. ~Discussions and Conclusions}

In the present study the WKB approximation technique is used to derive 
an alternate formalism for quantum systems with radial potentials,
confined within a rigid spherical box of radius $r_0$. Instead 
of the conventional Langer modification, in this approach the 
centrifugal term is decomposed perturbatively (in powers 
of $\hbar$) into 2 terms --- , 
the classical centrifugal potential and a quantum correction,
following the analysis of Hainz and Grabert [22].
The unique advantage of this approach is that it requires 
no modification of the centrifugal term in the WKB 
expansion when applied to radial potentials. Moreover, the 
quantization rules follow naturally from the 
WKB connection formulae, and the calculations are straightforward, though
somewhat lengthy. 

As a testing ground, the analysis is applied 
explicitly to 3 widely studied confined systems, each
of considerable importance in atomic phenomena, viz., 
the 3-dimensional Harmonic oscillator, the hydrogen atom 
and the Hulthen potential.
Each system is confined within a rigid spherical box of radius $r_0$. 
The spatial confinement imposes the additional boundary 
condition $~ \psi (r) ~=~ 0 ~ $ at $~r~=~r_0~$ on the radial wave
function. This criterion, alongwith the WKB connection formula, 
gives the quantization rules for estimating the energy eigenvalues, E.
For each case the results are computed and presented in tabular form, 
in Tables I - V, for ease of comparison with 
those obtained from other approximation  methods, viz.,
the conventional WKB method for confined systems 
(using the Langer modification of the
centrifugal term [18], variational results 
of Marin and Cruz [9], modified form of the same as given by 
Varshni [25] (in case of confined Hydrogen atom ), shifted 1/N expansion
method [15] (in case of the confined Hulth\'{e}n potential), 
and exact numerical results. 

It is easy to observe from the Tables I to V, that the present formalism
works quite well for all the 3 cases. 
The energy values are better than the conventional WKB energies 
(with Langer modification), as well as 
the shifted 1/N expansion results, justifying the perturbative expansion of
the centrifugal term even for potentials under hard-wall confinement. 
In some cases these energies are even 
better than the variational results of Marin and Cruz [9].
This is true for most of the confining radii, 
except when the size of the box is close 
to the turning point (marked by $^*$ in the tables). This is expected 
as the  WKB approximation is not valid close to the turning points. 
It may be worth mentioning here that Hainz and Grabert version of the WKB
method addresses primarily problems with singular potentials. Though the
harmonic oscillator does not fall in this class strictly, neverthelass, 
the centrifugal term
introduces a singularity at the origin for non-zero $l$. So the author found 
it interesting to check the behaviour of the confined harmonic
oscillator under such an expansion.

Another point worth examining is the effect of the higher order terms. 
It was shown in ref. [22] that in contrast to Langer modification, the higher
than first order terms gave a vanishing contribution to the estimate for
energy. However, this fact does not hold for confined potentials as is evident
below : \\
Expanding in powers of $\hbar$, one obtains
\beq
\int_{r_1} ^{r_0} \Gamma (r) dr  = \int_{r_1} ^{r_0} \Gamma _0 (r)dr 
- \int_{r_1} ^{r_0} \f {L_0}{2 \hbar \Gamma _0 r^2 }dr 
- \int_{r_1} ^{r_0} \f {L_0 ^2}{8 \hbar ^2  \Gamma _0 ^3 r^4 }dr 
- \int_{r_1} ^{r_0} \f {L_0 ^3}{16 \hbar ^3 \Gamma _0 ^5 r^6 }dr 
- \cdots
\eeq
Proceeding along analogous lines as above it was found that the eigenvalues
obtained in the lowest order became worse when higher order corrections were
evaluated. 

To conclude, this present analysis of perturbative decomposition 
of the centrifugal term into 2 parts ---
a classical potential and a quantum correction --- plays a vital role in
improving the WKB quantisation rule in the first order, 
thus yielding better estimates of the
energy eigenvalues. Hence this formalism may be useful in determining
the energy spectrum of any 3-dimensional radially confined problem.

\vs{1cm}

\section*{Acknowledgment}
  
The author is grateful to Prof. R. Roychoudhury for some useful discussions on
the topic. She thanks the referee for his / her invaluable comments, without
which the work could not have been presented in this form. 
She also acknowledges financial assistance from the Council 
of Scientific and Industrial Research, India.

\pb

\noindent
{\large \bf \un{Reference :}}

\singlespacing

\begin{enumerate}

\item[~1.] P. O. Fr\"{o}man, S. Yngve and N. Fr\"{o}man. J. Math. Phys. 
{\bf 28} (1987) 1813.
\item[~2.] C. Zicovich-Wilson, W. Jask\'{o}lski and J. H. Planelles.
Int. J. Quant. Chem.{\bf 54} (1995) 61,  {\it ibid.} {\bf 50} (1994) 429.
\item[~3.] C. Zicovich-Wilson, A. Corma and P. Viruela. J. Phys. Chem. 
{\bf 98} (1994) 10863.
\item[~4.] S.A. Cruz, E Ley-Koo, J.L. Marin and A. Taylor Armitage. Int. J.
Quan. Chem.  {\bf 54} (1995) 3.
\item[~5.] L. Jacak, P. Hawrylak and A. W\'{o}js. Quantum Dots.
{\it Springer} (1997).
\item[~6.] D.M. Larsen and S.Y. Mc Cann.
\item[] a) \ Phys. Rev. B {\bf 45} (1992) 3485-3488.
\item[] b) \ Phys. Rev. B {\bf 46} (1992) 3966-3970.
\item[~7] J.W. Brown and H.N. Spector.
\item[] a) \ J. Appl. Phys. (1986) {\bf 59} 1179-1180.
\item[] b) \ Phys. Rev. B (1987) {\bf 35} 3009-3012.
\item[~8.] A. Sinha and N. Nag. J. Math. Chem. {\bf 29} (2001) 267,
{\it and references therein}.
\item[~9.] J.L. Marin and S.A. Cruz. 
\item[] a) J. Phys. B : At. Mol. Opt. Phys. (1991) {\bf 24} 2899.
\item[] b) Am J Phys. {\bf 59} (1991) 931.
\item[10.] D. Keeports. Am. J. Phys. {\bf 58} (1990) 230.
\item[11.] M. J. El-Said. J. Phys. {\it I} France  {\bf 5} (1995) 1027.
\item[12.] M. N. Sergeenko. 
\item[] a) Phys. Rev. A {\bf 53} (1996) 3798.
\item[] b) Mod. Phys. Lett. A {\bf 13} (1998) 33.
\item[] c) Mod. Phys. Lett.  {\bf 12} (1997) 2859.
\item[] d) Mod. Phys. Lett. A {\bf 15} (2000) 83.
\item[13.] R. Vawter. Phys. Rev. {\bf 174} (1968) 749.

\pb

\item[14.] R. N. Kesarwani and Y.P.Varshni. 
\item[] a) J. Math. Phys.  {\bf 22} (1981) 1983.
\item[] b) J. Math. Phys.  {\bf 23} (1981) 803.
\item[15.] A. Sinha, R. Roychoudhury and Y. P. Varshni. Can. J. Phys. {\bf 78}
(2000) 141.
\item[16.] A. Sinha and R. Roychoudhury. Int. J. Quan. Chem. {\bf 73} (1999)
497. 
\item[17.] A. Sinha. Int. J. Quan. Chem. {\bf 79} (2000) 267.
\item[18.] A. Sinha, R. Roychoudhury and Y. P. Varshni. Can. J. Phys. 
Physica {\bf B 325} (2003) 214.
\item[19.] R. Langer. Phys. Rev. {\bf 51} (1937) 669.
\item[20.] I. H. Duru and H. Kleinert. Phys. Lett. {\bf B 84} (1979) 185.
\item[21.] R. Dutt, A. Khare and U. P. Sukhatme. Phys. Lett. {\bf B 181} (1986) 
295.
\item[22.] J. Hainz and H. Grabert. Phys. Rev. A {\bf 60} (1999) 1698.
\item[23.] A. K. Ghatak, R. L. Gallawa and I. C. Goyal. MAF and WKB Solutions
to the Wave Equations. NIST Monograph 176, Washington, DC, 1991.
\item[24.] I. S. Gradshteyn and I. M. Ryzhik. 
{\bf Tables of Integrals, Series and Products.} 
{\it Academic Press} (1992).
\item[25.] Y. P. Varshni. J. Phys. B : At. Mol. Opt. Phys. (1997)
{\bf 30} L589-L593, {\it and references therein}.
\item[26.] R. Dutt, A. Mukherjee and Y. P. Varshni. Phys. Rev. A {\bf 52}
(1995) 1750.

\end{enumerate}

\ed